\documentstyle[aps,prl,twocolumn]{revtex}   

\begin{document}

\wideabs{

\draft

\title{Adiabatic noise-induced escape rate for nonequilibrium open systems}

\author{Suman Kumar Banik, Jyotipratim Ray Chaudhuri and Deb Shankar 
Ray$^\star$}

\address{Department of Physical Chemistry,
Indian Association for the Cultivation of Science, Jadavpur,
Calcutta 700032, India.}

\date{\today}

\maketitle

\begin{abstract}
We consider the motion of an overdamped particle in a force field in presence 
of an external, adiabatic noise, without the restriction that the noise 
process is Gaussian or the stochastic process is Markovian. We examine the 
condition for attainment of steady state for this nonequilibrium open system 
and calculate the adiabatic noise-induced rate of escape of the particle over 
a barrier.
\end{abstract}


}

\narrowtext


\section{Introduction}

The motion of a Brownian particle in a fluid was first explained by Einstein 
in terms of fast thermal motion of fluid molecules striking the Brownian 
particle and causing it to undergo a random walk. One essential requirement
of the theory is that the noise is of {\it internal} origin. This implies
that the dissipative force which the Brownian particle experiences in course
of its motion in the fluid and the stochastic force acting on the particle as
a result of random impact of molecules arise from a common 
mechanism. From a microscopic point of view the system-reservoir Hamiltonian
description developed over the last few decades \cite{louisell,west} suggests that the coupling
of the system and the reservoir co-ordinates determines both the noise and
the dissipative terms of the Langevin equation describing the motion of the
particle. It is therefore not difficult
to anticipate that these two entities get related through the celebrated
fluctuation-dissipation theorem (Einstein's relation for diffusion and 
mobility is the first of its kind). The spiritual root of 
fluctuation-dissipation relation lies at the dynamical balance between inward 
flow of energy due to fluctuation of the reservoir into the system and the 
outward flow of energy from the system to the reservoir due to dissipation of 
the system. This dynamical balance is essential for attainment of a steady
state of the nonequilibrium system. In their treatise on nonequilibrium
statistical mechanics Lindenberg and West \cite{west} have classified these
systems as thermodynamically closed. However, there are quite a large number
of physical situations, a comprehensive account of which has been given in 
\cite{west}, where 
the system is thermodynamically open, i.e., when the system is driven by an 
{\it external} noise which is {\it independent} of system's characteristic
damping. These are important for describing the effects of pump fluctuations
on the emission of a dye laser \cite{roy}, effects of fluctuating rate constants on a
chemical reaction \cite{arnold}, 
effects of noise on electronic parametric oscillators \cite{kawa} etc.
The dynamics is still governed by a Langevin equation. Although there exists no 
relationship between the fluctuation and the dissipation in such a situation
it is interesting to search for the condition under which the steady state
of the thermodynamically open system is attained. Physically the sytem is
described by three timescales ; the timescale of dissipation or relaxation
$\beta^{-1}$ unlike the closed system is independent of correlation time
$\tau_c$ of the system. In the present
communication we consider the system to be driven by fluctuation which is
adiabatically slow such that it is characterized by a very long correlation 
time $\tau_c$ where
\begin{equation}
\label{ineqa}
\beta^{-1} \ll \Delta t \ll \tau_c \; \; .
\end{equation}

\noindent
Here $\Delta t$ refers to the timescale over which we look for the average
motion of the system. The latter inequality implies that $\beta^{-1}$, 
i.e., the inverse of damping constant defines the shortest timescale in the 
dynamics in contrast to the case of standard Brownian dynamics which obeys
$\beta^{-1} \gg \Delta t \gg \tau_c$. Secondly, we do not put the restriction 
that the stochastic processes is Markovian or the noise is Gaussian. The two 
assumptions have been discussed so much in the recent literaure that it is 
necessary to emphasize that the present consideration is free from these 
assumptions. We mention, in passing, that some interesting limitting 
situations have been examined for linearized systems driven by Gaussian
noise processes \cite{jaume1,jaume2}.

Our aim in the present article is two-fold. {\it First}, we examine the 
condition for a steady state distribution of the system driven by an external 
noise obeying the inequality (\ref{ineqa}). {\it Second}, Having obtained
this condition we calculate the {external, adiabatic} noise induced steady
state escape rate over a barrier in the spirit of Kramers'-Smoluchowski
theory. We show how the third-order noise plays an important role in both of 
these issues of stochastic dynamics.

\section{Third order noise and steady state probability density for open 
systems}

To start with we consider the 
equation of motion of a particle of unit mass in one dimension when
it is acted upon by an external field of force corresponding 
to a potential $V(x)$ and an {\it external, adiabatic} stochastic force 
$\xi(t)$,
\begin{equation}
\label{langevin}
\dot{x} = -\frac{1}{\beta} V'(x) + \frac{1}{\beta} \; \xi(t) \; \; .
\end{equation}

\noindent
where $\beta$ and the correlation time $\tau_c$ of $\xi(t)$ satisfy the 
inequality (\ref{ineqa}). Also note that by virtue of this we have considered 
the overdamped limit.
In a preceding paper \cite{jpa2} the equation of motion for probability 
density distribution function $P(x,t)$ in phase space corresponding to the 
Langevin description (\ref{langevin}) was derived. It has been shown that 
$P(x,t)$ obeys the differential equation of motion which contains third order 
terms (beyond the usual Fokker-Planck terms) giving rise to third order noise.
The appearance of these terms is generic for the stochastic process 
pertaining to the separation of timescales (\ref{ineqa}) we 
consider here. The general expression for time evolution of probability 
density function is given by \cite{jpa2}
\begin{eqnarray}
\label{fpeq}
\frac{\partial P(x,t)}{\partial t} & = & \frac{1}{\beta} 
\frac{\partial}{\partial x} \left [ V'(x) P(x,t) \right ] + 
\frac{c_{01}}{\beta^2} \frac{\partial^2 P(x,t)}{\partial x^2} \nonumber\\
& & -  \frac{c_2}{\beta^3} \frac{\partial^3}{\partial x^3}
\left [ V'(x) P(x,t) \right ] \; .
\end{eqnarray}

\noindent
$c_0$, $c_1$ and $c_2$ in Eq.(\ref{fpeq}) are as follows :
\begin{equation}
\left. \begin{array}{lll}
c_{01} & = & c_0 - c_1 \\
c_0 & = & \int_0^\infty \langle \xi(t) \; \xi(t-\tau) \rangle \; d\tau\\
c_1 & = & \int_0^\infty \tau \; \langle \xi(t) \; 
\left. \frac{ d\xi(t)}{dt} \right |_{(t-\tau)} \rangle \; d\tau \\
c_2 & = & \int_0^\infty \tau \; \langle \xi(t) \; \xi(t-\tau) \rangle 
\; d\tau \end{array} \right \} \; \; ,
\end{equation}

\noindent
where we have also assumed, for convenience $\langle \xi(t) \rangle = 0$.

The equation (\ref{fpeq}) describes the time evolution of an overdamped particle
in a force field (derivable from a potential $V(x)$) simultaneously subjected
to an external adiabatic stochastic force. $c_0$, $c_1$ and $c_2$ measure
the strength of the noise term. While the first term in 
Eq.(\ref{fpeq}) can be identified
as the usual deterministic dynamical term, the second and the third terms involve
second and third order diffusion coefficients due to stochasticity of $\xi(t)$.
The expansion scheme associated with Eq.(\ref{fpeq}) should not be confused
with Kramers-Moyal [KM] expansion (consideration of third order diffusion 
term of a KM expansion may lead to serious interpretive difficulties 
\cite{gardiner} because the probability distribution function often 
turns out to be negative) which serves as the standard starting point for 
analysis of stochastic processes with fast noise corresponding to the 
separation of timescale $\beta^{-1} \gg \Delta t \gg \tau_c$.
What is implicit in a KM expansion
is that the moments are {\it assumed} to be linear in $\tau$ and thus the
validity of the coefficients of a KM expansion rests on the smallness of
$\tau$. The cummulant expansion \cite{vanrep} similarly relies on the
smallness of correlation time $\tau_c$. On the other hand Eq.(\ref{fpeq})
is based on ``adibatic following approximation'' \cite{crisp,jpa1}
which involves an expansion in $1/\beta$ \cite{jpa2,jpa1}, which is evident 
from the nature of the coefficients of the terms in the right hand side.
The remarkable departure from the standard form of Fokker-Planck equation
(Smoluchowski equation) is due to the presence of the {\it third order noise}.
For other details we refer to \cite{jpa2,gardiner,garrido}.

In the next step we recast the third order equation (\ref{fpeq}) in the 
form of the familiar continuity equation and identify a steady state current 
$J$ in the following equation for a steady state probability distribution
function $P(x)$
\begin{equation}
\label{inhom}
\frac{d^2}{dx^2} \left \{ V'(x) P \right \} -a\frac{dP}{dx} -b\left \{
V'(x) P\right \} = \frac{\beta^3 J}{c_2} \; \; ,
\end{equation}

\noindent
where $a$ and $b$ are given by
\begin{equation}
\label{abval}
a = \frac{\beta c_{01}}{c_2} \; \; \; \; {\rm and} \; \; \; \;
b = \frac{\beta^2}{c_2} \; \; .
\end{equation}

\noindent
We consider a Kramers type potential $V(x)$ shown in Fig.(1) and
look for the steady state current at the barrier top by linearizing the 
potential at $x=0$. If $\omega_0$ refers to frequency of the inverted well 
and $E_0$ defines the potential at the barrier top, 
one obtains the following inhomogenous modified Bessel equation of order $\nu$
\begin{equation}
\label{mbf}
\zeta^2\frac{d^2 W}{d\zeta^2} + \zeta \; \frac{dW}{d\zeta} - 
(\nu^2 + \zeta^2 ) \; W = \frac{D}{b^{\frac{1}{2}(1+\nu)}}  \; 
\zeta^{1+\nu} \; \; .
\end{equation}

\noindent
In the above derivation following transformations and abbreviations have been
used
\begin{mathletters}
\begin{equation}
\label{trans1}
P(x) = x^{(1-\gamma)/2}\; W(x) \; \; , \; \; \zeta = \sqrt{b} x \; \; ,
\end{equation}

\begin{equation}
\label{trans2}
D = -\frac{\beta^3 \; J}{c_2 \; \omega_0^2} \; , \;
\gamma = \frac{2 \omega_0^2 + a}{\omega_0^2} \; , \;
\nu = \frac{1}{2} \left ( 1+ \frac{a}{\omega_0^2} \right ) \; \; .
\end{equation}
\end{mathletters}

\noindent
The homogenous counterpart corresponding to the above Eq.(\ref{mbf}) 
is the standard modified Bessel equation of order
$\nu$. The general solution of Eq.(\ref{mbf}) can be written as
\begin{eqnarray}
W(\zeta) & = &  A\; I_\nu(\zeta) + B\; K_\nu(\zeta) \nonumber \\
& & + \frac{D}{b^{\frac{1}{2}(1+\nu)}} \; I_\nu(\zeta) \int^\zeta \zeta^{\prime\nu}
\; K_\nu (\zeta') \; d\zeta' \nonumber \\
& & - \frac{D}{b^{\frac{1}{2}(1+\nu)}} \; K_\nu(\zeta) \int^\zeta \zeta^{\prime\nu}
\; I_\nu (\zeta') \; d\zeta' \; \; ,
\end{eqnarray}

\noindent
where $I_\nu(\zeta)$ and $K_\nu(\zeta)$ are modified Bessel functions of order
$\nu$; $A$ and $B$ are the two arbitrary constants of integration corresponding 
to the homogenous part of Eq.(\ref{mbf}). The $D$ containing term results from 
the particular integral corresponding to the inhomogenous contribution of 
Eq.(\ref{mbf}) obtained by the method of variation of parameters. 
Making use of the relations (\ref{trans1}) we revert back to the original 
variables $x$ and $P(x)$ to obtain the general solution of Eq.(\ref{mbf}) as
\begin{eqnarray}
\label{gsol}
P(x) & = & A \; x^{-\nu} \; I_\nu (\sqrt{b} x) +
B \; x^{-\nu} \; K_\nu (\sqrt{b} x)  \nonumber\\
& & +  D \; x^{-\nu}\; \left [ I_\nu (\sqrt{b} x) \int^{\sqrt{b}x} 
x^{'\nu} K_\nu (\sqrt{b} x') dx' \right. \nonumber \\
& & \left. - K_\nu (\sqrt{b} x) \int^{\sqrt{b}x} 
x^{'\nu} I_\nu (\sqrt{b} x') dx' \right ] \; \; .
\end{eqnarray}

We now impose two natural boundary conditions on the solution for the 
probability density function ;
(i) $P(x)$ vanishes for $|x|$ $\rightarrow$ $\infty$ and
(ii) $P(x)$ remains finite for all $x$.
The first condition ascribes the essential notion of a probability function 
which demands $A=0$. The second one is a necessary requirement since the 
relevant modified Bessel function is singular at $x=0$. This singularity at 
$x=0$ can be removed by constructing a relation between the constants $B$ 
and $D$ of the form 
\begin{equation}
B=(-1)^n D \frac{n!}{\sqrt{\pi}} 2^{ (2n+1)/2 } \frac{1}{ b^{(2n+3)/4} }
\; \; , 
\end{equation}

\noindent
such that all powers of $\frac{1}{x}$ vanishes 
identically from the exact solution \cite{bo} of the modified Bessel equation
(\ref{mbf}) which satisfy both the boundary conditions simultaneously provided
$\nu$ is an odd half integer, i.e.,
\begin{eqnarray*}
\nu = n + (1/2) \; \; , \; \; n = 1, 2, \ldots
\end{eqnarray*}

\noindent
so that from (\ref{abval}) and the definition of $\nu$ in (\ref{trans2}) we 
have (note that $n=0$ is not physically allowed)
\begin{equation}
\label{fdr}
\frac{\beta \; c_{01}}{2\; \omega_0^2 \; c_2} = n \; \; , \; \; 
n = 1, 2, \ldots \; \; .
\end{equation}

\noindent
The above Eq.(\ref{fdr}) relates the dissipation constant $\beta$ to the 
correlation of fluctuations ( $c_{01}$, $c_2$ ) of the external adiabatic 
noise as defined in Eq.(4). This is a fluctuation-dissipation like relation 
for the open systems. The integers $n$ characterize the distinct stable 
steady states in terms of the positive definite probability distribution 
function as given by (for $x>0$)
\begin{eqnarray}
\label{pdfn}
P_{n+\frac{1}{2}} (x) & = & (-1)^n \; D \; 2^n \; n! \; \frac{1}{b^{(n+2)/2}}
\sum_{k=0}^n f_k^n \; \frac{ e^{ -\sqrt{b} \; x} }{ x^{k+n+1} }
\nonumber\\
& & - \; \frac{D}{2\sqrt{b}} \sum_{i=0}^n \sum_{k=0}^n \sum_{j=0}^{n-k}
\{ (-1)^i + (-1)^{j-n} \} \nonumber \\
& & \times \; f_i^n \; f_k^n \; \frac{(n-k)!}{j!} \;
\frac{x^{j-i-n-1}}{ (\sqrt{b})^{n-k-j+1} }
\end{eqnarray}

\noindent
with
\begin{equation}
f_k^n = \frac{ (n+k)! }{ 2^k  b^{k/2} k! (n-k)! } \; \; . 
\end{equation}

The normalization constant $D$ is related to the steady state current $J$ 
(positive, implying a flow from left to right)
through the first of the relations (\ref{trans2}) and is given by
\begin{eqnarray}
\label{dnorm}
D & = & b \; \left [ 2  {\rm Ein}(\sqrt{b} \Delta) \right. \nonumber\\
& & \left. + 2^{n+1} n! \sum_{k=0}^n 
\sum_{j=0}^{k+n-1} \frac{(-1)^k}{2^k k! (n-k)! (n+k-j)}  \right ]^{-1}
\end{eqnarray}

\noindent
where $\Delta$ is large but finite and the function ${\rm Ein}(x)$ is defined 
\cite{as} as 
\begin{equation}
{\rm Ein}(x) = \sum_{k=1}^\infty (-1)^k \frac{x^k}{k! k} \; \; .
\end{equation}

\noindent
For $x<0$, the corresponding 
probability distribution function can be calculated from the symmetry
of the differential equation (\ref{mbf}). 

Expression (\ref{pdfn}) reduces to
the following simple form for, say, $n=1$, which is depicted in Fig.(2) for
several values of third order noise strength $c_2$
\begin{mathletters}
\begin{eqnarray}
P_{3/2}(x) & = & -2D \; b^{-3/2} \; e^{-\sqrt{b} x} \; \left ( \frac{1}{x^2}
+\frac{1}{\sqrt{b} x^3} \right ) \nonumber \\
& & - \frac{D}{b} \; \left ( \frac{1}{x} -
\frac{2}{bx^3} \right ) \; \; ; \; \; x\; >\; 0 \\
& = & -2D \; b^{-3/2} \; \; e^{\sqrt{b} x} \; \left ( \frac{1}{x^2}
- \frac{1}{\sqrt{b} x^3} \right ) \nonumber \\
& & + \frac{D}{b} \; \left ( \frac{1}{x} -
\frac{2}{bx^3} \right ) \; \; ; \; \; x\; <\; 0  \\
& = &  -\frac{ 2\; D}{ 3 \; \sqrt{b} } \; \; ; \; \; x \; = \; 0
\end{eqnarray}
\end{mathletters}

\noindent
Thus the steady states are physically allowed
only when one can relate the dissipation $\beta$ to the strength of second  
and third order external noise for these integers $n$.
Our analysis suggests that such a steady state condition (\ref{fdr})
for the open systems  can be realized at least in the specific issues as in 
the present instance. The third 
order noise is an essential ingredient for the dynamic balance we have 
referred to. It is thus also apparent why the thermodynamic open systems 
described by a Fokker-Planck equation for fast fluctuations which include only 
second order diffusion coefficient reaches a steady state with no 
fluctuation-dissipation relation.

\section{ Dynamics of barrier crossing induced by adiabatic noise }

Once the condition for attainment of the steady state is realized it becomes 
possible to consider the situation such that a particle in a force field, e.g.,
originally confined in a potential well may escape under the influence of
external adiabatic noise by maintaining a steady state current over the 
barrier. It is therefore pertinent to calculate the 
noise-induced escape rate in the spirit of Kramers and Smoluchowski and to 
elucidate the aspect of dependence of escape rate on dissipation. The 
counterpart of the latter issue in the theory of fast fluctuation is the 
wellknown turn-over problem in Kramers' theory \cite{kramers,rate}.

To proceed further we again make use of a Kramers' type potential $V(x)$ 
( Fig.(1) ) under 
which the particle moves, in the third order equation (\ref{fpeq}) for the 
probability distribution function. The popular flux-over-population method 
\cite{kramers,rate}
is then employed. The calculation rests on the evaluation of two quantities;
(i) the steady state current $J$ over the barrier top, (located at $x=0$) 
that results if the particles are continuously fed into the domain
of attraction (say, in the region of left well) and are subsequently and 
continuously removed in the neighboring domain of attraction. (ii) steady 
state population $n_a$ in the initial domain of attraction, i.e., the 
left well. The rate is defined by
\begin{equation}
\label{jbyn}
{\cal K} = J/n_a \; \; .
\end{equation}

For linearization the potential $V(x)$ at the bottom of the 
left well at $x= -\Delta$ we approximate
$V(x) \simeq \frac{1}{2} \; \omega_b^2 \; (x+\Delta)^2 $,
where $\omega_b$ refers to the frequency at the bottom of the left well. 
The above approximation to left well and $J=0$ condition reduce the third order 
equation of motion (Eq.(\ref{fpeq})) for probability density $P_b(x)$ near 
the bottom of the well to the following form. 
\begin{equation}
\label{bottom}
(x+\Delta) \; \frac{d^2 P_b}{dx^2} + \gamma' \; \frac{d P_b}{dx} -
b \; (x+\Delta) \; P_b = 0 \; \; .
\end{equation}

\noindent
The above equation is valid near the bottom of the left well 
($x \simeq -\Delta$). Here $\gamma'$ is defined as

\begin{equation}
\gamma' = \frac{2\omega_b^2 - a}{\omega_b^2} 
\end{equation}

\noindent
and $a$ by Eq.(\ref{abval}). Eq.(\ref{bottom}) can be written as
\begin{equation}
\label{bmbf}
z^2 \; \frac{d^2 W}{dz^2} + z \; \frac{d W}{dz} -
[\nu^{\prime 2}+ b \; z^2] \; W = 0 \; \; ,
\end{equation}

\noindent
making use of the following sets of of transformations,
\begin{mathletters}
\begin{equation}
\label{btrans1}
y(z) = z^{\frac{1}{2} (\sigma +1)} \; W(z) \; \; , \; \; 
\nu'=\frac{\sigma-1}{2} \; \; ,
\end{equation}

\begin{equation}
\label{btrans2}
2 - \gamma' = \sigma \; \; , \; \; z = x+\Delta \; \; , \; \; 
y = z P_b \; \; .
\end{equation}
\end{mathletters}

The solutions of Eq.(\ref{bmbf}) are again the modified Bessel functions. 
Reverting back to original variables, the general solution for the steady
state probability distribution near the bottom of the left well is given by
\begin{eqnarray}
\label{bgsol}
P_b(x) & = & A' \; (x+\Delta)^{\nu'} \; I_{\nu'} [\sqrt{b}\; (x+\Delta)] 
\nonumber \\
& & + B' \; (x+\Delta)^{\nu'} \; K_{\nu'} [\sqrt{b}\; (x+\Delta)] \; \; ,
\end{eqnarray}

\noindent
$A'$ and $B'$ are the two arbitrary constants of integration. 
The solution $P_b(x)$ must satisfy the following boundary conditions ;
(i) $P_b(x)$ must vanish at $x \rightarrow \infty$ and
(ii) $ P_b(-\Delta) = P_t^{n+\frac{1}{2}} (-\Delta)$,
where the stationary probability $P_t^{n+\frac{1}{2}} (-\Delta)$ corresponds
to the vanishing current $J=0$ along $x$ pertaining to the homogenous version
of Eq.(\ref{mbf}). As usual, $P_t^{n+\frac{1}{2}} (x)$  must also satisfy 
the boundary condition that for $|x| \rightarrow \infty$, 
$P_t^{n+\frac{1}{2}} (x)$ vanishes. The first condition leads to $A'=0$ and
the second one gives the value of $B'$ in terms of $B$ of Eq.(\ref{gsol})
and therefore
\begin{equation}
\label{pbsol}
P_b(x) = B' \; (x+\Delta)^{\nu'} \; K_{\nu'}[\sqrt{b} \; (x+\Delta)] \; \; .
\end{equation}

\noindent
where
\begin{eqnarray}
\label{bpval}
B' & = & B \; \sqrt{\frac{\pi}{2\; b^{1/2}}} \; 
\frac{b^{\nu'/2}}{\Gamma(\nu') \; 2^{\nu' -1}} \nonumber \\
& & \sum_{k=0}^n 
(-1)^{k+n+1} \; f_k^n \; \frac{e^{\sqrt{b} \Delta}}{\Delta^{k+n+1}} \; \; .
\end{eqnarray}

\noindent
Therefore $P_b(x)$ in Eq.(\ref{pbsol}) may be expressed as
\begin{eqnarray}
P_b(x) & = & B \; \sqrt{\frac{\pi}{2\; b^{1/2}}} \; 
\frac{b^{\nu'/2}}{\Gamma(\nu') \; 2^{\nu' -1}} \nonumber \\
& & \sum_{k=0}^n 
(-1)^{k+n+1} \; f_k^n \; \frac{e^{\sqrt{b} \Delta}}{\Delta^{k+n+1}} 
(x+\Delta)^{\nu'} \nonumber \\
& & \times \; K_{\nu'} [ \sqrt{b} (x+\Delta) ] \; .
\end{eqnarray}

\noindent
The above distribution which is valid near the bottom of the left well may be 
used to calculate the population inside the left well as,
\begin{equation}
\label{nanorm}
n_a = 2 \int_{-\Delta}^0 P_b(x) \; dx \; \; .
\end{equation}

\noindent
Due to the presence of $K_{\nu} (x)$ the probability $P_b(x)$ is a 
rapidly decreasing function. We may extend the above integration limit to
infinity. This yields ( using Eq.(\ref{pbsol}) ) \cite{gr}
\begin{equation}
\label{naval}
n_a = B' \; \frac{ \sqrt{\pi}\; 2^{\nu'} \; \Gamma(\nu' +\frac{1}{2}) }
{ (\sqrt{b} )^{\nu' +1} } \; \; .
\end{equation}

\noindent
Using the relations (\ref{bpval}) and (\ref{naval}) we finally have
\begin{eqnarray}
n_a & = & \sqrt{2}\; \pi \; B \; \frac{ \Gamma (\nu' +\frac{1}{2})}{\Gamma (\nu')} 
\nonumber \\
& & \sum_{k=0}^n (-1)^{k+n+1} \; f_k^n \; 
\frac{  e^{\sqrt{b} \Delta}}{\Delta^{k+n+1}} \; b^{-3/4} \; \; .
\end{eqnarray}

Having determined the population of the left well, $n_a$ and the steady state 
current, $J$ ( from first of the relations in (\ref{trans2}) and (\ref{dnorm})
) over the barrier we 
are now in a position to calculate the escape rate ${\cal K}_{n+\frac{1}{2}}$
($=J/n_a$). We quote the final result :
\begin{eqnarray}
{\cal K}_{n+\frac{1}{2}} & = &
\frac{c_2 \omega_0^2}{\sqrt{\pi} \beta^3} \; \frac{1}{n!} \; 
\frac{\Gamma(n \frac{\omega_0^2}{\omega_b^2} - \frac{1}{2} )}
{\Gamma(n \frac{\omega_0^2}{\omega_b^2} )} \nonumber \\
& & \left \{ \frac{e^{-\sqrt{b} \Delta} }{
\sum_{k=0}^n (-1)^k \; \frac{(n+k)!}{k! \; (n-k)!} \; 
\frac{2^{n-k+1}}{\Delta^{k+n+1}} \; \frac{1}{b^{(n+k+3)/2}} 
} \right \} .
\end{eqnarray}

\noindent
$\Delta$-s refer to the zero's of the potential $V(x)$ as shown in Fig.(1)
and by virtue of linearization of $V(x)$ at $x=0$ ;
$\Delta$ is approximately given by
$\Delta = ( 2 E_0/ \omega_0^2 )^{1/2}$ .

The above expression can be made more transparent by demonstrating a
representative transition rate, say, for $n=1$ as follows :
\begin{equation}
{\cal K}_{3/2} = \frac{1}{8\sqrt{\pi}} \frac{c_{01}}{c_2^2}
\frac{\Gamma (\frac{\beta c_{01}}{2 c_2 \omega_b^2} -\frac{1}{2} )}
{\Gamma (\frac{\beta c_{01}}{2 c_2  \omega_b^2} )} \;
(\Delta \beta)^2 \exp \left (-\frac{\beta \Delta}{\sqrt{c_2}} \right )
\end{equation}

The above expression is analogous to Kramers' formula for the rate of escape 
from a potential well over a finite barrier of height $E_0$ under the
influence of an external nonthermal adiabatic noise pertaining 
to the timescale (\ref{ineqa}).

The escape rate expressions derived above suggest that the rate approaches 
zero both for $\beta \rightarrow$ large and $\beta \rightarrow$ small. This 
behavior is somewhat reminiscent of Kramers' theory \cite{kramers}, where it 
was noted 
earlier that this two limiting behavior implies a maximal rate at some
damping value $\beta$. The rate therefore undergoes a turnover in a form of a
bell-shaped curve. In Fig.(3) we plot a representative variation of the
escape rate versus dissipation $\beta$ for different third order noise 
strength. With increasing friction, the rate undergoes a turnover from an
increasing behavior at low friction to an inverse behavior in the high 
friction limit. 

\section{Conclusion}

In conclusion, we consider the motion of particle in a force field, 
simultaneously subjected to adiabatic fluctuations of external origin. The 
equation of motion for probability distribution function includes a third 
order noise term. We 
show that although the system is thermodynamically open, the
specific interplay of the characteristic dissipation of the system and the 
correlation of fluctuations due to external 
non-Gaussian noise leads to distinct steady 
states for the open system. We calculate the external adiabatic
noise-induced rate of escape of a particle confined in a well. A typical 
variation of the escape rate as a function of dissipation which is 
reminiscent of Kramers' turn-over problem, has been demonstrated. In view
of several experimental investigations on external noise induced transitions
in the recent past \cite{west,roy,arnold,kawa}, 
the study of thermodynamically open systems has been
specially relevant in both physical and chemical sciences. Although the
driving noise processes in these cases are fast, suitable extension to
adiabatic noise limit may throw new light on the present issue. The 
population inversion in a two-level system by an adiabatically varying 
stochastic electromagnetic field, as considered earlier in 
Refs.\cite{jpa1,gris} may also be worthwhile candidate for further studies 
in this context.

\acknowledgments
SKB is indebted to Council of Scientific and Industrial Research (C.S.I.R.), 
Government of India for financial support. We express our sincerest thanks to 
Prof. J. K. Bhattacharjee for various discussions. DSR thanks 
Prof. N. Satyamurthy for his kind invitation for the present contribution.


\begin{center}
{\large{ Figure Captions } }
\end{center}

\begin{enumerate}

\item {\bf Fig.(1)} : A schematic plot of the Kramers' type potential $V(x)$ 
used for the calculation of barrier crossing dynamics.

\item {\bf Fig.(2)} : The normalized probability distribution function
$P_{3/2}(x)$ is plotted as a function of $x$ for various values of
the third order noise strength $c_2$ ( $\beta =1.0$ and $\Delta = 2.5$ ).

\item {\bf Fig.(3)} : Escape rate ${\cal K}_{3/2}$ is plotted as a function of 
the characteristic dissipation $\beta$ of the system for various values of 
$c_2$ ( $c_{01}=7.0$, $\omega_b = 0.80$ and $\Delta = 2.5$ ).

\end{enumerate}

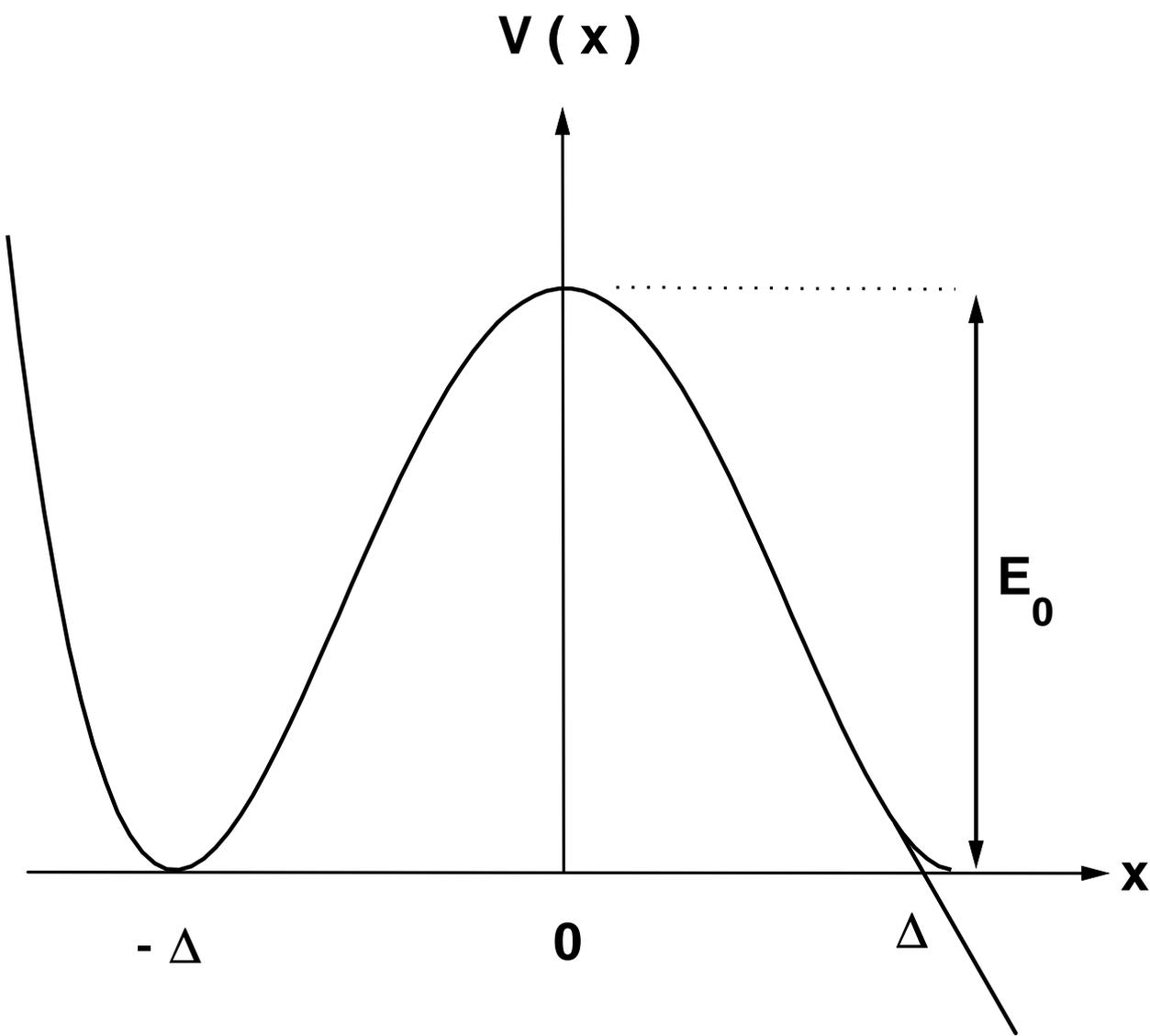
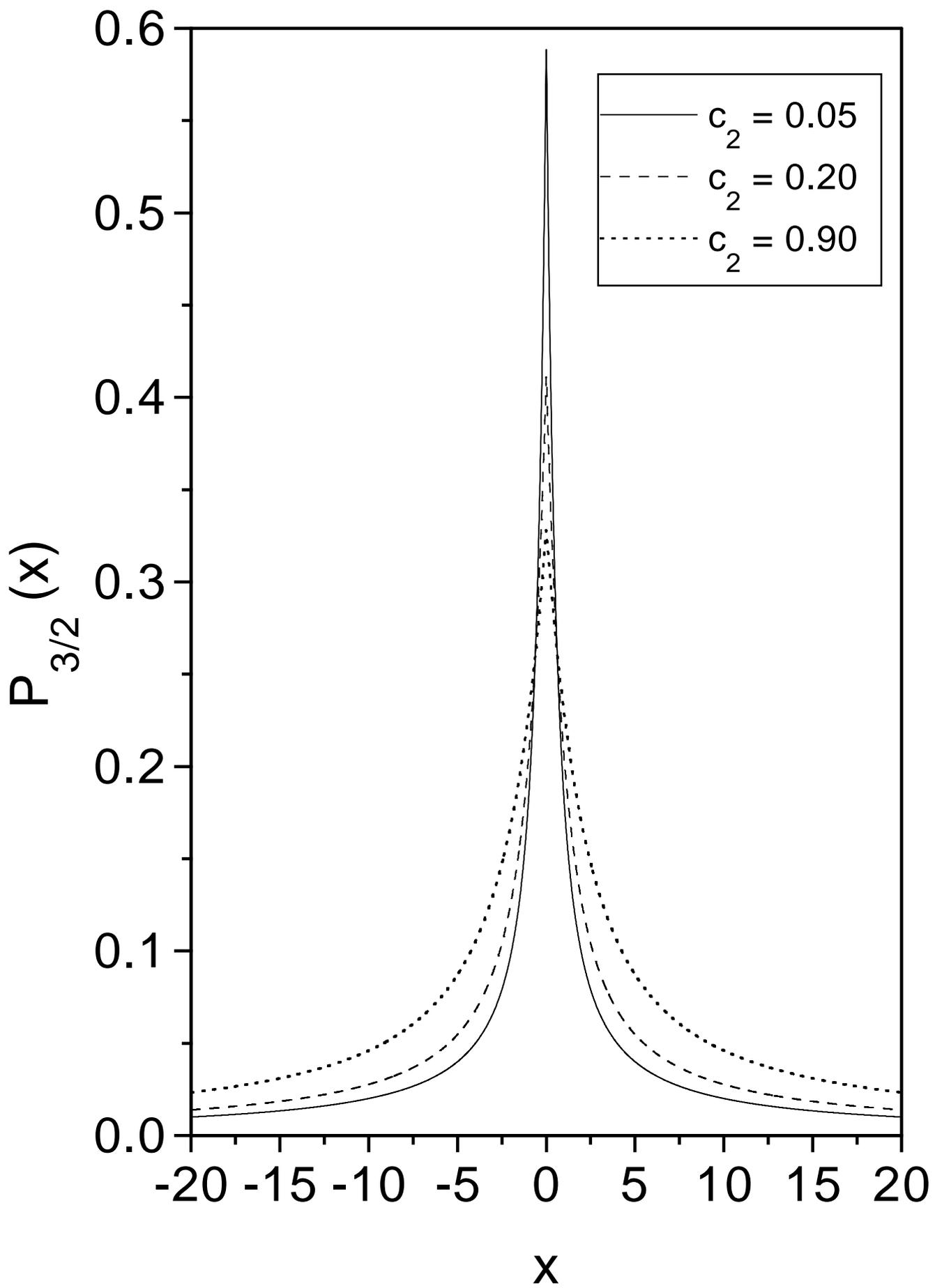
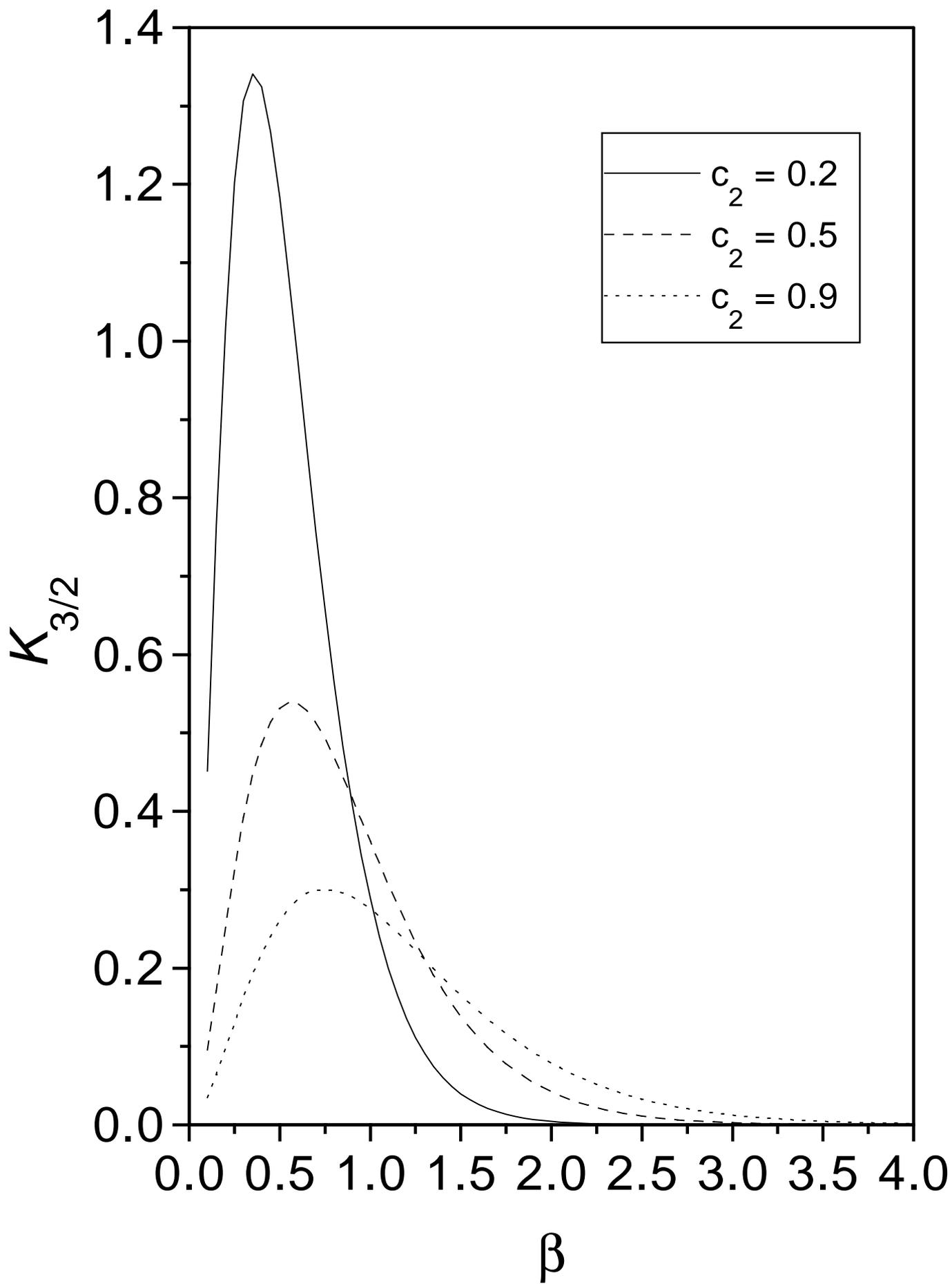
\end{document}